\documentclass[aps,prl,twocolumn,groupedaddress,showpacs]{revtex4}
\usepackage{graphicx}
\usepackage{amssymb}

\begin{document}


\title{What is in a pebble shape?}


\author{D.J. Durian,$^{1,2}$ H. Bideaud,$^{2}$ P. Duringer,$^{3}$ A.
Schr\"oder,$^{2}$ F. Thalmann,$^{2}$ C.M. Marques$^{2}$}
\affiliation{$^{1}$Department of Physics and Astronomy, University
of Pennsylvania, Philadelphia, PA 19104-6396, USA}
\affiliation{$^{2}$LDFC-CNRS UMR 7506, 3 rue de l'Universit\'e,
67084 Strasbourg Cedex, France} \affiliation{$^{3}$CGS-CNRS UMR
7517, Institut de G\'eologie, 1 rue Blessig, 67084 Strasbourg Cedex,
France}


\date{\today}

\begin{abstract}
We propose to characterize the shapes of flat pebbles in terms of
the statistical distribution of curvatures measured along the pebble
contour.  This is demonstrated for the erosion of clay pebbles in a
controlled laboratory apparatus. Photographs at various stages of
erosion are analyzed, and compared with two models.  We find that
the curvature distribution complements the usual measurement of
aspect ratio, and connects naturally to erosion processes that are
typically faster at protruding regions of high curvature.
\end{abstract}



\maketitle


In soft matter physics it is well understood why certain objects
have naturally rounded shapes. For example, liquid droplets are
round due to surface tension~\cite{adamson} and bilayer vesicles are
round due to bending rigidity~\cite{lipowsky}. In both cases the
equilibrium shapes are determined by free-energy minimization, and
sinusoidal perturbations decay exponentially with time according to
a power of the wavevector.  But how do pebbles in a river or on a
beach come to have their familiar, smooth, rounded
shapes~\cite{aristotle}? Are there similarly simple physical
principles that govern the evolution of freshly-fractured polyhedral
rocks into smooth rounded pebbles? Aristotle proposed that erosion
is more rapid at regions farther from the center, where greater
impulses can be more readily delivered~\cite{aristotle}. However
this idea has not been quantified or tested, and we are aware of no
other models for the origin of pebble shapes.

A wealth of knowledge already exists on naturally-occurring
pebbles~\cite{boggs}, thanks to decades of field and laboratory
study~\cite{wentworth19, wadell32, krumbein41a, rayleigh, sneed,
wald}. There is a rich terminology for describing sizes (grain,
pebble, cobble, boulder) and shapes (angular, rounded, elongated,
platy). The simplest quantitative shape description is to form
dimensionless indices from the lengths of long vs intermediate vs
short axes. There is still debate as to how these axes are best
defined, as well as to which dimensionless ratios are most
useful~\cite{illenberger, benn, howard, hofmann, graham}. Another
quantitative method used more for grains than for flat pebbles
involves Fourier transform of the contour~\cite{boggs, ehrlich,
clark, thomas}. There the relative amplitudes of different harmonics
give an indication of shape in terms of roughness at different
length scales.  However a Fourier description is not particularly
natural because peaks erode faster than valleys, so that sinusoidal
perturbations cannot erode without changing shape. To make progress
in understanding the origin of pebble shapes, one needs a
quantitative shape description in terms of microscopic variables
directly relevant to the erosion process.

In this paper we propose to quantify the shape of flat pebbles in
terms of {\it curvature}, measured at each point along the entire
two-dimensional contour. The curvature $K$ equals the reciprocal
radius of a circle that locally matches the contour, and can be
deduced from the coordinates of the pebble
boundary~\cite{weisstein}.  This could be useful, since one can
imagine faster erosion at protruding regions of higher positive
curvature, and perhaps no erosion for scalloped regions of negative
curvature.  To illustrate the use of curvature for shape
description, we perform a laboratory experiment in which clay
pebbles are eroded from controlled initial shapes.  We find that the
curvature distribution along the pebble contour becomes stationary,
indicating a final fixed-point shape. Surprisingly, this stationary
shape is not a perfect circle. We then introduce two models of
pebble erosion, and test them using the curvature distribution as
well as the aspect ratio. We find that the curvature distribution is
a more incisive tool for discriminating between models.  These
results point the way for future studies, both of
naturally-occurring pebbles and of lab-eroded pebbles of more direct
geophysical interest.

\begin{figure}
\includegraphics[width=2.80in]{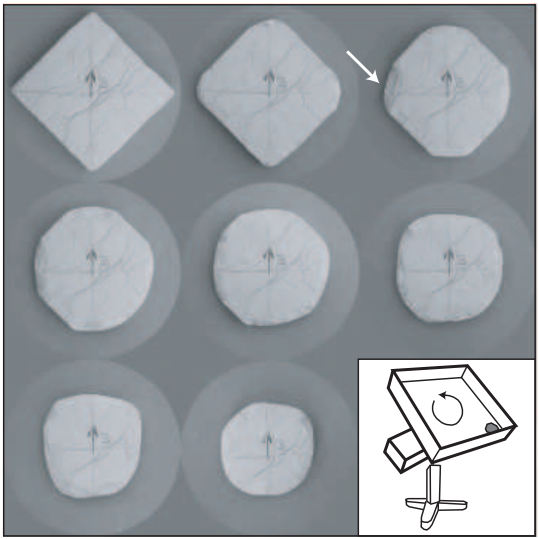}
\caption{\label{Carre3} Shape evolution of a
$5\times5\times0.5$~cm$^3$ square pebble eroded in a rotating basin,
depicted on the lower right, with roughly 300 cutting events between
each photograph. The time sequence is left to right, and top to
bottom.  The erosion is due to small random cutting events, a
particularly large example of which is highlighted by arrow in the
third photograph.}
\end{figure}

\begin{figure}
\includegraphics[width=2.80in]{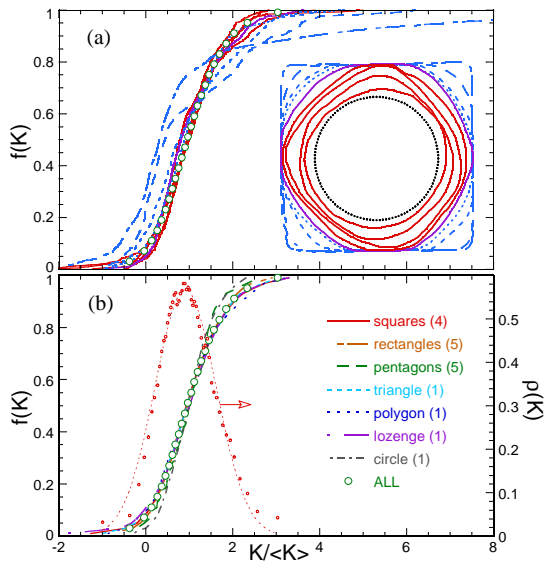}
\caption{\label{Lab} [Color online] (a) Digitized contours (inset)
and cumulative curvature distribution for the initially-square clay
pebble of Fig.~\protect{\ref{Carre3}}.  Note that the red curves are
indistinguishable, indicating a stationary final shape. The final
shape is not circular, as demonstrated both by the black-dotted
circle in the inset and by the gradual rise of $f(K)$ from zero to
one.  (b) Left axis: Cumulative curvature distribution for the final
shapes attained by clay pebbles of varied initial shape, as labeled;
the number of runs for each shape is given in parentheses. Right
axis: average curvature distribution for all final shapes, $\rho=\rm
df/\rm dK$, and best fit to a Gaussian.}
\end{figure}

For our laboratory experiments, the pebbles were prepared from
ordinary white clay (``chamotte''), molded into different polygonal
shapes with uniform $0.5$~cm thickness. Individual pebbles were
eroded one at a time in a square pan, $30\times 30$~cm$^2$ with 7~cm
walls, spun at 1~Hz around the central axis perpendicular to the
bottom surface and oriented at $45^\circ$ away from vertical. The
pebble is first dragged upwards, at rest on the pan; near the top it
begins to slide, and it accelerates until striking the wall; then it
rolls along the side, comes to rest, and starts a new cycle. Erosion
is mainly caused by collision with the walls.  At five minute
intervals, corresponding to roughly 300 collision events, the clay
pebble is removed and photographed. As an example, photographs of
one initially-square pebble with sides of 5~cm are shown in
Fig.~\ref{Carre3}; digitized contours are shown in the inset of
Fig.~\ref{Lab}a. Evidently the erosion is fastest at the corners,
which protrude and have high positive curvature. Once the corners
have been removed, the pebble reaches a nearly round shape that
progressively shrinks.

The first step in analysis is to deduce the value of the curvature
at each digitized point along the contour by linear-least-squares
fits of radius vs angle to a third-order polynomial.  Some care is
needed in choice of fitting window: if it is too small then the
matrix inversion of the fit is singular~\cite{nr}, if it is too
large then the fit deviates systematically from the contour points;
results for all acceptable windows are averaged together.  As a
check, the perimeter $P$ is computed from straight-line segments
between adjacent points and is confirmed to equal $2\pi$ divided by
the average curvature $\langle K\rangle$~\cite{weisstein}.

As a shape-descriptor, we next construct the curvature distribution
such that $\rho(K)\rm dK$ is the probability of finding a value of
the curvature between $K$ and $K+\rm dK$.  For reliability, it is
preferable to work with the cumulative curvature distribution,
$f(K)=\int_0^K\rho(K')\rm dK'$, which may be computed as the
fraction of the perimeter with curvature less than $K$~\cite{nr}.
For a perfect circle, $\rho(K)$ is a delta-function and $f(K)$ is a
step-function. The evolution of $f(K)$ for one square pebble is
shown in the main plot of Fig.~\ref{Lab}a, corresponding to the
contours of the inset. Note that the results are plotted vs
$K/\langle K\rangle$ in order to remove a trivial scale factor
related to the ever-shrinking perimeter.  Evidently, the curvature
distribution starts broad but narrows down as erosion proceeds. By
about the fifth (purple) or sixth (largest red) contour, the form of
$f(K)$ vs $K/\langle K\rangle$ reaches a steady state that
fluctuates around a well-defined average.  If the initial shape is
circular, then by contrast the form of $f(K)$ starts narrow but
broadens as erosion proceeds.

To test the possible influence of initial conditions on the final
shape, we repeat the erosion experiment of Fig.~\ref{Lab}a for clay
pebbles molded into several different shapes.  In all cases the form
of $f(K)$ vs $K/\langle K\rangle$ eventually becomes time
independent, with results given for different initial shapes on the
left axis of Fig.~\ref{Lab}b. To within fluctuations comparable with
those seen for a single stationary pebble as it shrinks further,
these are all identical.  Thus the final shape is truly {\it
stationary}, independent of both size and the initial conditions,
and may be quantified by averaging together the final cumulative
curvature distributions for all pebbles; the results are shown by
open circles in Fig.~\ref{Lab}b.  Note that its form is not set by
uncertainty in curvature measurement, since the circular pebble
started with a narrower distribution.

The stationary curvature distribution, obtained by numerical
differentiation of the average $f(K)$, is shown on the right axis of
Fig.~\ref{Lab}b.  Clearly it is not a delta-function, and hence the
shape is not a perfect circle.  The form of $\rho(K)$ can be
reasonably fit to a Gaussian, as shown, with a standard deviation of
$\sigma/\langle K\rangle=0.7$. However, the actual distribution is
skewed in the tails and has a standard deviation of $\sigma/\langle
K\rangle=0.8$. Though our laboratory erosion experiment may be
artificial from a geophysical perspective, it is interesting that a
stationary shape exists and that it is not trivially circular.
Clearly, $f(K)$ was a useful tool for this demonstration.

\begin{figure}
\includegraphics[width=2.80in]{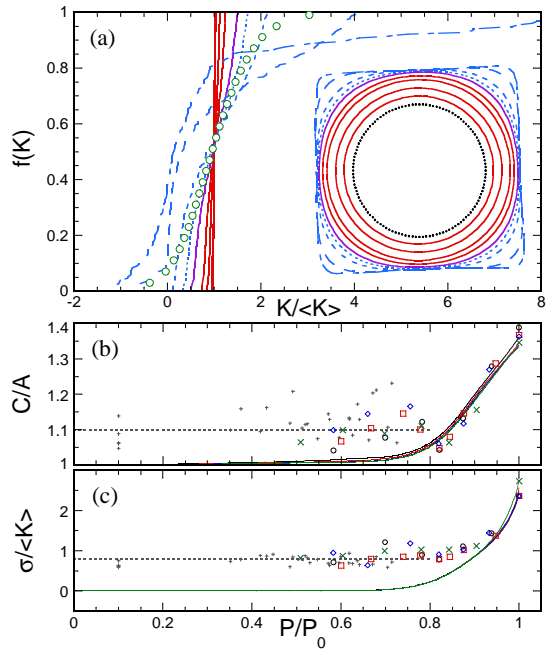}
\caption{\label{Polish} [Color online] (a) Contours (inset) and
cumulative curvature distribution predicted by the {\it ``polishing
model''} for a square pebble.  The initial contour points, the
sequence of perimeters displayed, and the line-color codes, are the
same as in Fig.~\protect{\ref{Lab}}a. The open green circles
represent the stationary $f(K)$ averaged over all laboratory
pebbles.  (b) The caliper aspect ratio and (c) the standard
deviation of the curvature distribution vs scaled perimeter; note
that time evolution is from right to left.  Symbol and line codes
for (a) and (b) are as follows: Data for four different square
pebbles are shown by large colored symbols; data for pebbles of
other initial shape are shown as small gray pluses once stationarity
has been reached; for the five pentagons $P_\circ$ was not measured,
so data are shown arbitrarily at $P/P_\circ=0.1$; aspect ratio and
standard deviation for the final stationary shape of the laboratory
experiments are shown by dashed lines. Predictions of the polishing
model, applied to the four square pebbles, are shown by solid
curves.}
\end{figure}

In the remainder of the paper, we use the cumulative curvature
distribution to evaluate the validity of two simple models of the
laboratory erosion experiment.  We begin with a {\it ``polishing
model''} in which erosion is directly coupled to curvature. Namely,
the normal velocity at each boundary point is taken in proportion to
curvature, for regions of positive curvature, and is zero otherwise.
(If all points are moved normally in proportion to curvature,
irrespective of sign, then it has been proven that any initial
contour will shrink to a point and approach a circle in this
limit~\cite{grayson}.)  The evolution predicted by the polishing
model for a square pebble is shown in Fig.~\ref{Polish}.  In the top
plot, Fig.~\ref{Polish}a, the contours and cumulative curvature
distributions are plotted at the same sequence of perimeters, and
with the same line-color codes, as the actual data in
Fig.~\ref{Lab}a.  As expected, this model is highly efficient at
polishing a rough pebble into a smooth shape. It becomes essentially
circular, to the eye, soon after the entire original boundary has
been eroded.  The cumulative curvature distribution, $f(K)$, appears
to approach a step function, contrary to our laboratory experiments
shown in Fig.~\ref{Lab}. Thus the polishing model may be appropriate
for other natural or laboratory erosion processes, but not for ours.

To track evolution toward a final shape more clearly, and to compare
new vs old shape description methods, the aspect ratio of the
pebbles and the width of the curvature distributions are shown in
Figs.~\ref{Polish}b-c. These are plotted vs perimeter divided by the
initial perimeter, as a surrogate for the duration of erosion
(except that time evolution is from right to left). Since major and
minor orthogonal axes are equal for a square shape, we work with the
``caliper'' aspect ratio, $C/A$, of largest to smallest separations
for parallel planes that confine the pebble; as the name suggests,
this quantity may be readily measured in the field using calipers.
For a dimensionless measure of the width of the curvature
distribution, we take the standard deviation, $\sigma$, divided by
the average curvature, $\langle K\rangle$.  The results for all
pebbles are plotted in Figs.~\ref{Polish}b-c, with four
initially-square pebbles highlighted by larger symbols. Stationarity
is reached when the square pebbles are eroded to about 3/4 of their
initial perimeter, after which both $C/A$ and $\sigma/\langle
K\rangle$ data fluctuate significantly about an average value. These
fluctuations represent real shape changes, not measurement
uncertainty, and are larger for $C/A$ than for $\sigma/\langle
K\rangle$. Predictions by the polishing model for the evolution of
the four square pebbles are also included for completeness. While
the initial agreement is satisfactory, the polishing predictions for
$C/A$ and $\sigma/\langle K\rangle$ quickly fall below the data and,
furthermore, do not exhibit fluctuations.

To more successfully explain the laboratory erosion, we introduce a
minimal one-parameter {\it ``cutting model''}.  We imagine the
effect of collision between pebble and wall is to create a straight
fracture near the boundary whose size, on average, is set by the
collision impulse; e.g. note the unusually large cut in the third
photograph in Fig.~\ref{Carre3}. At each erosion step of the model,
we thus pick a contour point at random and we choose a cut length
from an exponential distribution whose average is some fraction
$\alpha$ of the square-root of the pebble area. One can imagine
other choices, but an exponential is simple and the impulse is set
by the area rather than the perimeter.

\begin{figure}
\includegraphics[width=2.80in]{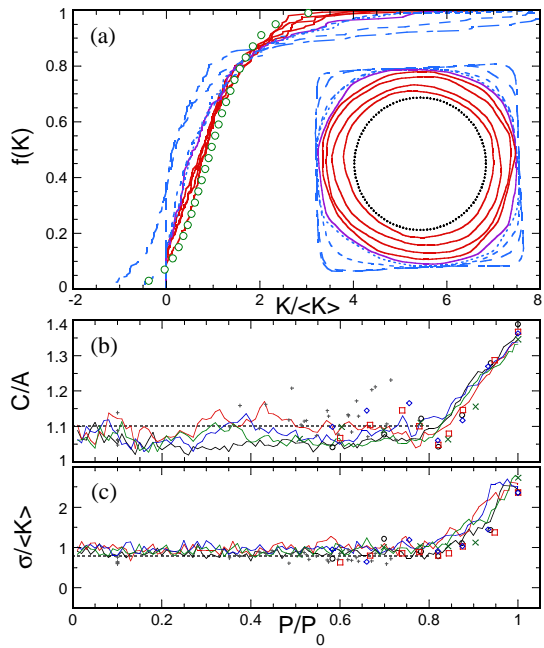}
\caption{\label{Cut}  [Color online] Predictions for evolution based
on the {\it ``cutting model''}, and comparison with data.  Displayed
quantities are the same as for the {\it ``polishing model''} in
Fig.~\protect{\ref{Polish}}.}
\end{figure}

To match the cutting model with our data, we take $\alpha=0.042$ in
order that the final, stationary shape have the same curvature
distribution width as the laboratory pebble data. The resulting
evolution of the contours and curvature distributions predicted for
one square pebble are shown in Fig.~\ref{Cut}a. As observed in the
lab, the corners erode faster than the flat sections and the final
shape is nearly, but not perfectly, circular.  The cutting model
predictions are compared directly with the $C/A$ and $\sigma/\langle
K\rangle$ data in Figs.~\ref{Cut}b-c. While the final average value
of $\sigma/\langle K\rangle$ is correct by design, our model also
captures several other features of the data: the final average value
of $C/A$, the magnitude of shape fluctuations in the stationary
shape, and the rate of evolution from initial to final shapes. In
spite of this success, the full form of $f(K)$ in Fig.~\ref{Cut}a is
not completely correct in fine detail. Particularly, the model does
not generate regions of negative curvature.  No doubt this could be
remedied by taking cuts from an appropriate distribution of shapes,
not just straight.  Such refinements could be tested by comparison
with $f(K)$ data.

In conclusion, the distribution of curvature along the contour of a
pebble is more incisive than the aspect ratio as a tool for both
describing pebble shapes and for testing models.  The
characterization of such eroded forms is of obvious importance in
the subfield of geology known as sedimentology~\cite{boggs}, where
pebbles are considered as witness to the geological conditions under
which they were formed. It would therefore be of great practical
consequence if there existed a method to decipher the message
imprinted in a pebble shape, a method that would distinguish beach
vs glacier vs river erosion, that would tell how far has a pebble
traveled down a stream and perhaps even for how long it has been
subjected to erosion forces. Moreover, the ability to quantify and
explain shapes resulting from abrasion, erosion or other wearing
mechanisms would also be of significance in other fields, wherever
nature or man transforms a solid object by irreversibly removing
fractions of the mass in a sequence of elementary events.  The
curvature distribution could prove a helpful tool to study all such
processes.

\begin{acknowledgments}
We acknowledge insightful discussions P. Boltenhagen.  This work was
supported by the Chemistry Department of the CNRS, under AIP
``Soutien aux Jeunes Equipes'' (CM).  It was also supported by the
National Science Foundation under Grant DMR-0514705 (DJD).
\end{acknowledgments}

\bibliography{PebRefs}

\begin{thebibliography}{21}
\expandafter\ifx\csname natexlab\endcsname\relax\def\natexlab#1{#1}\fi
\expandafter\ifx\csname bibnamefont\endcsname\relax
  \def\bibnamefont#1{#1}\fi
\expandafter\ifx\csname bibfnamefont\endcsname\relax
  \def\bibfnamefont#1{#1}\fi
\expandafter\ifx\csname citenamefont\endcsname\relax
  \def\citenamefont#1{#1}\fi
\expandafter\ifx\csname url\endcsname\relax
  \def\url#1{\texttt{#1}}\fi
\expandafter\ifx\csname urlprefix\endcsname\relax\def\urlprefix{URL }\fi
\providecommand{\bibinfo}[2]{#2}
\providecommand{\eprint}[2][]{\url{#2}}

\bibitem[{\citenamefont{Adamson}(1990)}]{adamson}
\bibinfo{author}{\bibfnamefont{A.}~\bibnamefont{Adamson}},
  \emph{\bibinfo{title}{Physical Chemistry of Surfaces}}
  (\bibinfo{publisher}{Wiley}, \bibinfo{address}{New York},
  \bibinfo{year}{1990}), \bibinfo{edition}{5th} ed.

\bibitem[{\citenamefont{Lipowsky and Sackmann}(1995)}]{lipowsky}
\bibinfo{author}{\bibfnamefont{R.}~\bibnamefont{Lipowsky}} \bibnamefont{and}
  \bibinfo{author}{\bibfnamefont{E.}~\bibnamefont{Sackmann}},
  \emph{\bibinfo{title}{Structure and Dynamics of Membranes}}
  (\bibinfo{publisher}{Elsevier}, \bibinfo{address}{Amsterdam},
  \bibinfo{year}{1995}).

\bibitem[{\citenamefont{Aristotle}(2000)}]{aristotle}
\bibinfo{author}{\bibnamefont{Aristotle}}, \emph{\bibinfo{title}{Minor Works,
  Mechanical Problems, Question 15}} (\bibinfo{publisher}{Harvard},
  \bibinfo{address}{Cambridge}, \bibinfo{year}{2000}), \bibinfo{note}{trans. by
  W.S. Hett.}

\bibitem[{\citenamefont{{S. Boggs, Jr.}}(2001)}]{boggs}
\bibinfo{author}{\bibnamefont{{S. Boggs, Jr.}}},
  \emph{\bibinfo{title}{Principles of Sedimentology and Stratigraphy}}
  (\bibinfo{publisher}{Prentice Hall}, \bibinfo{address}{N.J.},
  \bibinfo{year}{2001}), \bibinfo{edition}{3rd} ed.

\bibitem[{\citenamefont{Wentworth}(1919)}]{wentworth19}
\bibinfo{author}{\bibfnamefont{C.~K.} \bibnamefont{Wentworth}},
  \bibinfo{journal}{J. Geology} \textbf{\bibinfo{volume}{27}},
  \bibinfo{pages}{507} (\bibinfo{year}{1919}).

\bibitem[{\citenamefont{Wadell}(1932)}]{wadell32}
\bibinfo{author}{\bibfnamefont{H.}~\bibnamefont{Wadell}}, \bibinfo{journal}{J.
  Geology} \textbf{\bibinfo{volume}{40}}, \bibinfo{pages}{443}
  (\bibinfo{year}{1932}).

\bibitem[{\citenamefont{Krumbein}(1941)}]{krumbein41a}
\bibinfo{author}{\bibfnamefont{W.~C.} \bibnamefont{Krumbein}},
  \bibinfo{journal}{J. Sed. Pet.} \textbf{\bibinfo{volume}{11}},
  \bibinfo{pages}{64} (\bibinfo{year}{1941}).

\bibitem[{\citenamefont{{Lord Rayleigh, F.R.S.}}(1944)}]{rayleigh}
\bibinfo{author}{\bibnamefont{{Lord Rayleigh, F.R.S.}}},
  \bibinfo{journal}{Nature} \textbf{\bibinfo{volume}{3901}},
  \bibinfo{pages}{169} (\bibinfo{year}{1944}).

\bibitem[{\citenamefont{Sneed and Folk}(1958)}]{sneed}
\bibinfo{author}{\bibfnamefont{E.~D.} \bibnamefont{Sneed}} \bibnamefont{and}
  \bibinfo{author}{\bibfnamefont{R.~L.} \bibnamefont{Folk}},
  \bibinfo{journal}{J. Geology} \textbf{\bibinfo{volume}{66}},
  \bibinfo{pages}{114} (\bibinfo{year}{1958}).

\bibitem[{\citenamefont{Wald}(1990)}]{wald}
\bibinfo{author}{\bibfnamefont{Q.~R.} \bibnamefont{Wald}},
  \bibinfo{journal}{Nature} \textbf{\bibinfo{volume}{345}},
  \bibinfo{pages}{221} (\bibinfo{year}{1990}).

\bibitem[{\citenamefont{Illenberger}(1991)}]{illenberger}
\bibinfo{author}{\bibfnamefont{W.~K.} \bibnamefont{Illenberger}},
  \bibinfo{journal}{J. Sed. Pet.} \textbf{\bibinfo{volume}{61}},
  \bibinfo{pages}{756} (\bibinfo{year}{1991}).

\bibitem[{\citenamefont{Benn and Ballantyne}(1992)}]{benn}
\bibinfo{author}{\bibfnamefont{D.~I.} \bibnamefont{Benn}} \bibnamefont{and}
  \bibinfo{author}{\bibfnamefont{C.~K.} \bibnamefont{Ballantyne}},
  \bibinfo{journal}{J. Sed. Pet.} \textbf{\bibinfo{volume}{62}},
  \bibinfo{pages}{1147} (\bibinfo{year}{1992}).

\bibitem[{\citenamefont{Howard}(1992)}]{howard}
\bibinfo{author}{\bibfnamefont{J.~L.} \bibnamefont{Howard}},
  \bibinfo{journal}{Sedimentology} \textbf{\bibinfo{volume}{39}},
  \bibinfo{pages}{471} (\bibinfo{year}{1992}).

\bibitem[{\citenamefont{Hofmann}(1994)}]{hofmann}
\bibinfo{author}{\bibfnamefont{H.~J.} \bibnamefont{Hofmann}},
  \bibinfo{journal}{J. Sed. Res.} \textbf{\bibinfo{volume}{64}},
  \bibinfo{pages}{916} (\bibinfo{year}{1994}).

\bibitem[{\citenamefont{Graham and Midgley}(2000)}]{graham}
\bibinfo{author}{\bibfnamefont{D.~J.} \bibnamefont{Graham}} \bibnamefont{and}
  \bibinfo{author}{\bibfnamefont{N.~G.} \bibnamefont{Midgley}},
  \bibinfo{journal}{Earth Surface Processes and Landforms}
  \textbf{\bibinfo{volume}{25}}, \bibinfo{pages}{1473} (\bibinfo{year}{2000}).

\bibitem[{\citenamefont{Ehrlich and Weinberg}(1970)}]{ehrlich}
\bibinfo{author}{\bibfnamefont{R.}~\bibnamefont{Ehrlich}} \bibnamefont{and}
  \bibinfo{author}{\bibfnamefont{B.}~\bibnamefont{Weinberg}},
  \bibinfo{journal}{J. Sed. Pet.} \textbf{\bibinfo{volume}{40}},
  \bibinfo{pages}{205} (\bibinfo{year}{1970}).

\bibitem[{\citenamefont{Clark}(1981)}]{clark}
\bibinfo{author}{\bibfnamefont{M.~W.} \bibnamefont{Clark}},
  \bibinfo{journal}{J. Int. Assc. Math. Geo.} \textbf{\bibinfo{volume}{13}},
  \bibinfo{pages}{303} (\bibinfo{year}{1981}).

\bibitem[{\citenamefont{Thomas et~al.}(1995)\citenamefont{Thomas, Wiltshire,
  and Williams}}]{thomas}
\bibinfo{author}{\bibfnamefont{M.~C.} \bibnamefont{Thomas}},
  \bibinfo{author}{\bibfnamefont{R.~J.} \bibnamefont{Wiltshire}},
  \bibnamefont{and} \bibinfo{author}{\bibfnamefont{A.~T.}
  \bibnamefont{Williams}}, \bibinfo{journal}{Sedimentology}
  \textbf{\bibinfo{volume}{42}}, \bibinfo{pages}{635} (\bibinfo{year}{1995}).

\bibitem[{\citenamefont{Weisstein}(1999)}]{weisstein}
\bibinfo{author}{\bibfnamefont{E.~W.} \bibnamefont{Weisstein}},
  \emph{\bibinfo{title}{The CRC Concise Encyclopedia of Mathematics}}
  (\bibinfo{publisher}{CRC Press}, \bibinfo{address}{New York},
  \bibinfo{year}{1999}).

\bibitem[{\citenamefont{Press et~al.}(1992)\citenamefont{Press, Flannery,
  Teukolsky, and Vetterling}}]{nr}
\bibinfo{author}{\bibfnamefont{W.~H.} \bibnamefont{Press}},
  \bibinfo{author}{\bibfnamefont{B.~P.} \bibnamefont{Flannery}},
  \bibinfo{author}{\bibfnamefont{S.~A.} \bibnamefont{Teukolsky}},
  \bibnamefont{and} \bibinfo{author}{\bibfnamefont{W.~T.}
  \bibnamefont{Vetterling}}, \emph{\bibinfo{title}{Numerical Recipes in C}}
  (\bibinfo{publisher}{Cambridge Univ. Press}, \bibinfo{address}{NY},
  \bibinfo{year}{1992}), \bibinfo{edition}{2nd} ed.

\bibitem[{\citenamefont{Grayson}(1987)}]{grayson}
\bibinfo{author}{\bibfnamefont{M.~A.} \bibnamefont{Grayson}},
  \bibinfo{journal}{J. Diff. Geometry} \textbf{\bibinfo{volume}{26}},
  \bibinfo{pages}{285} (\bibinfo{year}{1987}).

\end{thebibliography}

\end{document}